\documentclass[twocolumn,reprint,amsmath,amssymb, aps,prl]{revtex4-2}
\usepackage[T1]{fontenc}
\usepackage[latin9]{inputenc}
\usepackage{color}
\usepackage{float}

\makeatletter

\usepackage{amsfonts,amssymb}
\usepackage{latexsym}
\usepackage{epsfig}
\usepackage{graphicx}
\usepackage{tikz}
\definecolor{greatblue}{RGB}{40,120,181}
\definecolor{greatred}{RGB}{200,36,35}
\usepackage[colorlinks,linkcolor=greatblue,anchorcolor=blue,citecolor=greatred]{hyperref}

\usepackage{graphicx}
\usepackage{dcolumn}
\usepackage{bm}
\usepackage{upgreek}
\usepackage{mathrsfs}

\newcommand{\nn}{\nonumber}
\newcommand{\ii}{{\rm i}}
\newcommand{\ep}{{\rm e}}
\newcommand{\Eif}{{\hat\upphi}{}}
\newcommand{\dZ}{{\odot}}

\newcommand{\rd}{{\rm d}}

\newcommand{\cM}{{\mathcal M}}
\newcommand{\cA}{{\mathcal A}}
\newcommand{\cD}{{\mathcal D}}
\newcommand{\cG}{{\mathcal G}}
\newcommand{\cL}{{\mathcal L}}

\newcommand{\ha}{{\hat a}}
\newcommand{\hb}{{\hat b}}
\newcommand{\hc}{{\hat c}}

\newcommand{\bmsigma}{{\bm\sigma}}
\newcommand{\bmh}{{\bm h}}
\newcommand{\bmA}{{\bm A}}

\newcommand{\Fwedge}{{\bigwedge}}
\newcommand{\Fdelta}{{\updelta}}
\newcommand{\FD}{{\mathscr{D}}}
\newcommand{\FA}{{\mathscr{A}}}
\newcommand{\FF}{{\mathscr{F}}}
\newcommand{\FU}{{\mathfrak{U}}}
\newcommand{\Fu}{{\mathscr{U}}}
\newcommand{\FG}{{\mathscr{G}}}

\makeatother

\usepackage[english]{babel}

\begin{document}

\title{A toy model for $p$-form gauge symmetry }

\author{Yi Yan$^{1}$}  

\author{Zhao-Long Wang$^{1,2,3}$}
\email{zlwang@nwu.edu.cn}
\affiliation{$^1$Institute of Modern Physics, Northwest University, Xi'an 710127, China
\\$^2$Peng Huanwu Center for Fundamental Theory, Xi'an 710127, China
\\$^3$Shaanxi Key Laboratory for Theoretical Physics Frontiers, Xi'an 710127, China}


\begin{abstract}
The abelian $(p+1)$-form gauge field is inherently coupled to the $p$-brane worldvolume. After quantization, the corresponding $p$-form gauge transformation is associated with the local phase ambiguity of the $p$-brane wave functional. In essence, the $p$-form gauge symmetry can be realized as a special construction of the generic 0-form gauge symmetry in the functional space of $p$-brane configurations. The non-abelian generalization is straightforward in the functional space language. To simplify the analysis, we further introduce a toy model where the infinite dimensional functional space of $p$-brane configurations is replaced by a finite dimensional matrix space.  After taking the symmetric trace in the matrix model, the original discussions of the $p$-form gauge symmetry can be inherited by the toy model.
\end{abstract}

\maketitle
\onecolumngrid

\section{Introduction}
The charged $p$-brane is inherently coupled to the abelian $(p+1)$-form gauge field $A_{[p+1]}$ through the term 
\begin{eqnarray}
Q_p\int_{V_{p+1}} A_{[p+1]}
\end{eqnarray}
in the worldvolume action. This system remains invariant under the $U(1)$ $p$-form gauge transformation
\begin{eqnarray}
\tilde A_{[p+1]}=A_{[p+1]}-\rd \Lambda_{[p]}\,.
\end{eqnarray} 
When $p=0$, it characterizes the interaction between charged particles and the electromagnetic field. The concept of $0$-form gauge symmetry has been extended to non-abelian groups \cite{Yang:1954ek} and plays a central role in modern physics.  For $p>0$, it is equally natural to consider the non-abelian extension of higher form gauge symmetries. For instance, the non-abelian $2$-form gauge field should appear in the worldvolume theory of $M_5$ brane \cite{Callan:1991ky,Kaplan:1995cp,Becker:1996my}. In the literature, various attempts\cite{Aganagic:1997zq, Hofman:2002ey, Schreiber:2004ma, Gustavsson:2008dy, Saemann:2010cp, Papageorgakis:2011xg, Breen:2001ie, Attal:2002ix, Aschieri:2003mw, Baez:2010ya, Rey:2010uz, Nepomechie:1982rb, Lambert:2010wm, Douglas:2010iu, Lambert:2010iw, Ho:2011ni, Chu:2011fd,Chu:2012um} have been made in constructing the non-abelian higher form gauge theory. However, the generic theory of non-abelian $p$-form gauge symmetry has yet to be established. In this paper, we start with analyzing the $p$-form gauge symmetry from the standard $p$-brane point of view, and then construct a toy model for $p$-branes which would simplify the construction of the target space action.   

\section{$p$-brane}
\vspace{2ex}
\noindent \underline{\bf{$p$-brane action}}
\vspace{2ex}

Let us consider a $p$-brane moving in a target space $\cM$ with coordinates $x^{M}$. The tension of the $p$-brane is $T_p=l_p^{-(p+1)}$, and its charge is taken to be $-Q_p$. The embedding of the $n=p+1$ dimensional worldvolume $V_{n}$ into $\cM$ is given by $x^{M}=X^{M}(\bmsigma)$, where $\bm\sigma^{\mu}=\{\tau,\sigma^i\}$ are the world volume coordinates. The $p$-brane naturally couples to the target space metric $G_{MN}(x)$ as well as the $(p+1)$-form gauge field 
\begin{eqnarray}
A_{[p+1]}(x)=\frac1{(p+1)!}A_{M_0\cdots M_p}(x)\rd x^{M_0}\wedge\cdots\wedge \rd x^{M_p}\,.
\end{eqnarray}
The induced quantities on $V_{n}$ are
\begin{eqnarray}
\bmh_{\mu\nu}&=&G_{MN}(X)\partial_{\mu}X^{M}\partial_{\nu}X^{N}\,,
\\
\bmA_{\mu_0\cdots\mu_p}&=&A_{M_0\cdots M_p}(X)\partial_{\mu_0}X^{M_0}\cdots\partial_{\mu_p}X^{M_p}\,.
\end{eqnarray}
The dynamics of the $p$-brane is governed by the action
\begin{eqnarray}\label{S_p}
S_p=\int_{V_n}\rd^n\bm\sigma\left[-T_p\sqrt{\bmh}
-\frac{Q_p}{(p+1)!}\varepsilon^{\mu_0\cdots\mu_p}\bmA_{\mu_0\cdots\mu_p}\right] 
\end{eqnarray}
where $\varepsilon^{\mu_0\cdots\mu_p}$ is total antisymmetric with $\varepsilon^{0\cdots p}=1$ and 
\begin{eqnarray}
\bmh=-\det(\bmh_{\mu\nu})\,.
\end{eqnarray}

The conjugation momentum density of this system is given by
\begin{eqnarray}
\Pi_{M}&=&\frac{\Fdelta S}{\Fdelta\dot X^{M}}=T_p P_{M}-Q_p\cA_{M}\,,
\end{eqnarray}
where $P_{M}$ is the mechanical momentum density
\begin{eqnarray}
P_{M}&=&-\bmh^{\frac12}\bmh^{0\mu}G_{MN}(X)\partial_{\mu}X^{N}\,,
\end{eqnarray}
and the additional term $\cA_{M}$  
\begin{eqnarray}
\cA_{M}&=&\frac1{p!}\varepsilon^{i_1\cdots i_p}A_{MM_1\cdots M_p}(X)\partial_{ i_1}X^{M_1}\cdots\partial_{ i_p}X^{M_p}\,,~~~~~~\varepsilon^{i_1\cdots i_p}\equiv\varepsilon^{0 i_1\cdots i_p}\,,
\end{eqnarray}
arises due to  the coupling with the $(p+1)$-form gauge field.  
The classical dynamics of the $p$-brane are entirely determined by the constraints of the mechanical momentum  
\begin{eqnarray}\label{HC}
G^{MN}(X)P_MP_{N}&=&-\det(\bmh_{ij})\,,~~~~~~~~~~~~~
P_{M}\partial_{i}X^{M}=0\,.
\end{eqnarray}

\vspace{2ex}
\noindent \underline{\bf{$p$-brane wave functional}}
\vspace{2ex}

At the off-shell level, this system can be canonically quantized by mapping the Poisson brackets to the equal-time commutators of the operators
\begin{eqnarray}\label{CQ}
&&[X^{M}(0,\sigma),\Pi_{N}(0,\tilde\sigma)]=\ii\delta^{M}_{N}\delta^{p}(\sigma-\tilde\sigma)\,,
\cr&&[ X^{M}(0,\sigma),X^{N}(0,\tilde\sigma)]=0\,,~~~~~~~~~~~ [\Pi_{M}(0,\sigma),\Pi_{N}(0,\tilde\sigma)]=0\,.
\end{eqnarray}
These operators act on the $p$-brane wave functional 
\begin{eqnarray}
\Psi[X^M(\sigma)]=\langle X^{M}(\sigma)|\Psi\rangle 
\end{eqnarray}
which is defined on the infinite dimensional space $\{X^{M}(\sigma)\}$ of $p$-brane equal time configurations. 
In the path integral formalism, the $p$-brane wave functional is expressed as 
\begin{eqnarray}
\Psi[X(\sigma)] &=&\int^{\tilde X(0,\sigma)=X(\sigma)}_{\Psi}[\cD \tilde X]\ep^{\ii S_p[\tilde X]}\,.
\end{eqnarray}
The canonical commutation relations (\ref{CQ}) imply that we can identify the canonical momentum density operator with the functional derivative operator  
\begin{eqnarray}
\Pi_{M}(\sigma)=-\ii \frac{\Fdelta}{\Fdelta X^{M}(\sigma)}\,.
\end{eqnarray}
Thus, the quantum operator for $\ii T_p P_{M}$ is the functional covariant derivative 
\begin{eqnarray}
\ii T_p P_{M}=\FD_{M}(\sigma)
= \frac{\Fdelta}{\Fdelta X^{M}(\sigma)}+\ii Q_p\cA_p\,.
\end{eqnarray}
The wave functional $\Psi[X^M(\sigma)]$ must satisfies the wave equations which are the operator equations for the Hamiltonian constraints (\ref{HC}).

\section{$p$-form gauge symmetry}
\subsection{Abelian $p$-form gauge symmetry}
\vspace{2ex}
\noindent \underline{\bf{Functional $U(1)$ gauge symmetry}}
\vspace{2ex}

When $p=0$, the local gauge symmetry is associated with the phase ambiguity of the corresponding wave functions of point particles. To discuss the higher form gauge symmetry, it is natural to consider the local $U(1)$ phase ambiguity of the $p$-brane wave functional $\Psi[X^M(\sigma)]$. Generically, it is 
\begin{eqnarray}\label{ApT}
\Psi[X^M(\sigma)]\sim \tilde\Psi[X^M(\sigma)]=\ep^{\ii Q_p\Theta[X^M(\sigma)]}\Psi[X^M(\sigma)]
\end{eqnarray}
where the $U(1)$ phase $\Theta[X^M(\sigma)]$ itself is a local functional in the $p$-brane configuration space $\{X^{M}(\sigma)\}$. It is generically non-local in the target space point of view.     
For simplicity, we set $Q_p=1$ in the following discussions. 

As in the $p=0$ case, one can introduce the corresponding functional gauge covariant derivative 
\begin{eqnarray}
\FD_{M}(\sigma)= \frac{\Fdelta}{\Fdelta X^{M}(\sigma)}+\ii \FA_{M}(\sigma)[X]\,.
\end{eqnarray}
The functional $U(1)$ gauge field 1-form in the $p$-brane configuration space $\{X^{M}(\sigma)\}$ is
\begin{eqnarray}
\FA[X]&=& \int_{\Sigma_p}\rd^p\sigma\,\Fdelta X^{M}(\sigma)\FA_{M}(\sigma)[X]\,,
\end{eqnarray}
where $\{\Fdelta X^{M}(\sigma)\}$ forms a basis of differential forms on $\{X^{M}(\sigma)\}$.  
Together with the total derivative operator 
\begin{eqnarray}
\Fdelta&=& \int_{\Sigma_p}\rd^p\sigma\,\Fdelta X^{M}(\sigma)\frac{\Fdelta}{\Fdelta X^{M}(\sigma)}\,,
\end{eqnarray}
in $\{X^{M}(\sigma)\}$, we have
\begin{eqnarray}
\FD&=&\int_{\Sigma_p}\rd^p\sigma\,\Fdelta X^{M}(\sigma)\FD_{M}(\sigma)=\Fdelta+\ii\FA[X]\,
\,.
\end{eqnarray}

The gauge covariance of $\FD$ demands that
\begin{eqnarray}
&&
\ep^{\ii\Theta[X]}\FD\Psi[X]=\tilde\FD\tilde\Psi[X]
\,.
\end{eqnarray}
It follows that the functional 1-form $U(1)$ gauge field must transform as
\begin{eqnarray}\label{AbpTr}
\tilde\FA[X]&=&
\FA[X]-\Fdelta\Theta[X]\,.
\end{eqnarray}
The commutator of the covariant derivatives gives rise to the functional field strength 2-form  
\begin{eqnarray}
\FF[X]&=&-\frac{\ii}2 \left[\FD,\FD\right]
=\frac12\int_{\Sigma_p}\rd^{p}\sigma_1\int_{\Sigma_p}\rd^{p}\sigma_2\,\Fdelta X^{M_1}(\sigma_1)\Fwedge\Fdelta X^{M_2}(\sigma_2)\FF_{M_1,M_2}(\sigma_1,\sigma_2)[X]\,,
\end{eqnarray}
where  $\Fwedge$ is the wedge product in the $p$-brane configuration space  $\{X^{M}(\sigma)\}$, and
\begin{eqnarray}
\FF_{M_1,M_2}(\sigma_1,\sigma_2)[X]&=&\frac{\Fdelta}{\Fdelta X^{M_1}(\sigma_1)}\FA_{M_2}(\sigma_2)[X]
-\frac{\Fdelta}{\Fdelta X^{M_2}(\sigma_2)}\FA_{M_1}(\sigma_1)[X]\,.
\end{eqnarray}
As in the $p=0$ case, $\FF[X]$ is invariant under the functional $U(1)$ gauge transformation. 

\vspace{2ex}
\noindent \underline{\bf{Abelian $p$-form gauge symmetry}}
\vspace{2ex}

Under the target space abelian $p$-form gauge transformation, the $(p+1)$-form gauge field transforms as
\begin{eqnarray}\label{Abp-Tr}
A_{[p+1]}(x)\to \tilde A_{[p+1]}(x)=A_{[p+1]}(x)-\rd\Lambda_{[p]}(x)\,.
\end{eqnarray} 
It induces a boundary term in the worldvolume action (\ref{S_p})
\begin{eqnarray}
\tilde S_{p}=S_{p}+\int_{V_n}\rd\Lambda_{[p]}=S_{p}+\int_{\partial V_n} \Lambda_{[p]}\,.
\end{eqnarray} 
Consequently, the $p$-brane wave functional 
\begin{eqnarray}
\Psi[X(\sigma)] &=&\int^{\tilde X(0,\sigma)=X(\sigma)}_{\Psi}[\cD \tilde X]\ep^{\ii S_p[\tilde X]}\,,
\end{eqnarray}
transforms as 
\begin{eqnarray}
\tilde\Psi[X(\sigma)]=\ep^{\ii \int_{\Sigma_p=\partial V_n} \Lambda_{[p]}}\Psi[X(\sigma)]\,.
\end{eqnarray}
Comparing with (\ref{ApT}), it is obvious that the $p$-form gauge symmetry is just the special realization 
\begin{eqnarray}
\Theta[X]&=&\int_{\Sigma_p}\Lambda_{[p]}(X) 
=\frac{1}{p!}\int_{\Sigma_p}\rd^p\sigma\,\varepsilon^{ i_1\cdots i_p}\Lambda_{M_1\cdots M_p}(X(\sigma))\partial_{ i_1}X^{M_1}(\sigma)\cdots\partial_{ i_p}X^{M_p}(\sigma)
\end{eqnarray}
of the local $U(1)$ transformation in $p$-brane configuration space $\{X^{M}(\sigma)\}$.

The corresponding functional 1-form gauge field in $\{X^{M}(\sigma)\}$ is 
{\fontsize{10 pt}{\baselineskip}\selectfont\begin{eqnarray}
\FA[X]&=&\frac{1}{p! }\int_{\Sigma_p}\rd^p\sigma\,\Fdelta X^{M}(\sigma)\varepsilon^{i_1\cdots i_p}A_{MM_1\cdots M_p}(X)\partial_{ i_1}X^{M_1}(\sigma)\cdots\partial_{ i_p}X^{M_p}(\sigma) 
=\int_{\Sigma_p} \iota_{\Fdelta X^{M}(\sigma)}A_{[p+1]}(X)\,. 
\end{eqnarray}
By using integration by parts on $\Sigma_p$, one can verify that the generic transformation rule (\ref{AbpTr}) correctly reproduces the initial $(p+1)$-form gauge field transformation rule  (\ref{Abp-Tr}) in this special realization
\begin{eqnarray}
\tilde\FA[X]&=&\FA[X]-\Fdelta\Theta[X]
=\int_{\Sigma_p}\iota_{\Fdelta X^{M}(\sigma)}\left[A_{[p+1]}(X)-\rd \Lambda_{[p]}(X)\right]\,.
\end{eqnarray}

Furthermore, it is natural to expect that the functional 2-form gauge field strength $\FF[X]$ is related to the $(p+2)$-form field strength  
\begin{eqnarray}
F_{[p+2]}=\rd A_{[p+1]}\,.
\end{eqnarray} 
In fact, given the boundary condition $\Fdelta X(\sigma)|_{\partial\Sigma_p}=0$,
we find that 
\begin{eqnarray}
\FF[X]
&=&\frac{1}{2\,p!}\int_{\Sigma_p}\rd^{p}\sigma\,
\Fdelta X^{M}(\sigma) \Fwedge \Fdelta X^{N}(\sigma)\varepsilon^{i_1\cdots i_p}F_{M NM_1\cdots M_p}(X)\partial_{ i_1}X^{M_1}(\sigma) \cdots\partial_{ i_p}X^{M_p}(\sigma)
\cr&=&\frac12\int_{\Sigma_p}\,   \iota_{\Fdelta X^{M}(\sigma)\Fwedge \Fdelta X^{N}(\sigma)}F_{[p+2]}(X(\sigma))
\,.
\end{eqnarray}

\subsection{Non-abelian generalization}
\vspace{2ex}
\noindent \underline{\bf{Functional non-abelian gauge symmetry}}
\vspace{2ex}

As in the $p=0$ case, we can further consider the wave functional multiplet $\Phi^{\hat A}[X(\sigma)]$ which forms a linear representation of a non-abelian group $\cG$ with the Lie algebra 
\begin{eqnarray}
[t_\ha,t_\hb]=\ii f^{\hc}{}{}_{\ha \hb}t_\hc\,.
\end{eqnarray}
At the infinitesimal level, the functional gauge transformation is 
\begin{eqnarray}
\FU[X]&=&1+\ii \Fu^\ha[X](t_\ha)^{\hat A}{}_{\hat B}\,. 
\end{eqnarray}
Correspondingly, we have the functional 1-form gauge field  
\begin{eqnarray}
\FA[X]&=&\FA^\ha[X](t_\ha)^{\hat A}{}_{\hat B}= \int_{\Sigma_p}\rd^p\sigma\,\Fdelta X^{M}(\sigma)\FA^{\ha}_{M}(\sigma)[X](t_\ha)^{\hat A}{}_{\hat B}\,,
\end{eqnarray}
as well as the functional gauge covariant derivative similar to the abelian case 
\begin{eqnarray}
\FD&=&\Fdelta+\ii\FA[X]
\,.
\end{eqnarray}
The gauge covariance of the functional covariant derivative
\begin{eqnarray}
&&\FU[X]\FD\Phi[X]=\tilde\FD (\FU[X]\Phi[X])\,, 
\end{eqnarray}
implies that
\begin{eqnarray}
\tilde\FA[X]=\FU[X]\FA[X]\FU^{-1}[X]-\Fdelta \FU[X] \FU^{-1}[X]\,.
\end{eqnarray}
At the infinitesimal level, it becomes 
\begin{eqnarray}
\delta_{\cG}\FA[X]=-\FD\Fu^\ha[X]\,. 
\end{eqnarray}

The functional field strength 2-form is 
\begin{eqnarray}
\FF[X]&=&
-\frac{\ii}2 \left[\FD,\FD\right]
=\frac12\int_{\Sigma_p}\rd^{p}\sigma_1\int_{\Sigma_p}\rd^{p}\sigma_2\,\Fdelta X^{M_1}(\sigma_1)\Fwedge\Fdelta X^{M_2}(\sigma_2)\FF^\ha_{M_1,M_2}(\sigma_1,\sigma_2)[X]t_\ha\,,~~~~~ 
\end{eqnarray}
where
\begin{eqnarray}
\FF^\ha_{M_1,M_2}(\sigma_1,\sigma_2)[X]&=&\frac{\Fdelta}{\Fdelta X^{M_1}(\sigma_1)}\FA^\ha_{M_2}(\sigma_2)[X]
-\frac{\Fdelta}{\Fdelta X^{M_2}(\sigma_2)}\FA^\ha_{M_1}(\sigma_1)[X]
+\ii f^{\ha}{}_{\hb_1\hb_2}\FA^{\hb_1}_{M_1}(\sigma_1)[X]\FA^{\hb_2}_{M_2}(\sigma_2)[X]\,. 
\end{eqnarray}

\vspace{2ex}
\noindent \underline{\bf{Non-abelian $p$-form gauge symmetry}}
\vspace{2ex}

Inspired by the abelian case, we can try to realize $\FA^\ha[X]$ specifically through a target space non-abelian $p$-form gauge field $A^\ha_{MM_1\cdots M_p}(X)$ as follows
\begin{eqnarray}
\FA^\ha[X]&=&\frac{1}{p! }\int_{\Sigma_p}\rd^p\sigma\,\Fdelta X^{M}(\sigma)\varepsilon^{i_1\cdots i_p}A^\ha_{MM_1\cdots M_p}(X)\partial_{ i_1}X^{M_1}(\sigma)\cdots\partial_{ i_p}X^{M_p}(\sigma) 
\cr&=&\int_{\Sigma_p}\iota_{\Fdelta X^{M}(\sigma)}A^\ha_{[p+1]}[X]\,.
\end{eqnarray}
The corresponding infinitesimal functional gauge transformation is expected to be 
\begin{eqnarray}
\Fu^\ha[X]&=&\int_{\Sigma_p} u^\ha_{[p]}[X]=\frac{1}{p! }\int_{\Sigma_p}\rd^p\sigma\, \varepsilon^{i_1\cdots i_p}u^\ha_{M_1\cdots M_p}(X)\partial_{ i_1}X^{M_1}(\sigma)\cdots\partial_{ i_p}X^{M_p}(\sigma) \,,
\end{eqnarray}
which is associated with a target space Lie-algebra valued $p$-form $u^\ha_{[p]}(x)$.

The infinitesimal functional gauge transformation of $\FA^\ha[X]$ then becomes
\begin{eqnarray}
\delta_{\cG}\FA^\ha[X]&=&-\FD\Fu^\ha[x]
=-\int_{\Sigma_p}\iota_{\Fdelta X^{M}(\sigma)} \rd u^\ha_{[p]}-\ii f^\ha{}_{\hb\hc} \int_{\Sigma_p}\iota_{\Fdelta X^{M}(\sigma)}\, A^\hb_{[p+1]}\int_{\Sigma_p}u^\hc_{[p]}\,. 
\end{eqnarray}
It implies that the target space non-abelian $p$-form gauge transformation of $A^\ha_{[p+1]}$ is   
\begin{eqnarray}
\delta_{\cG} A^\ha_{[p+1]}&=&-\rd u^\ha_{[p]}-\ii f^\ha{}_{\hb\hc} A^\hb_{[p+1]}\int_{\Sigma_p}u^\hc_{[p]}\,.
\end{eqnarray}
Although $\delta_{\cG}\FA^\ha[X]$ is local in the $p$-brane configuration space $\{X^{M}(\sigma)\}$, the induced $\delta_{\cG} A^\ha_{[p+1]}(x)$ is non-local from the target space point of view.

Under the present special realization, the functional 2-form field strength is  
\begin{eqnarray}
\FF^\ha[X]&=&\!\frac12\int_{\Sigma_p}\rd^{p}\sigma_1\int_{\Sigma_p}\rd^{p}\sigma_2\,\Fdelta X^{M_1}(\sigma_1)\Fwedge\Fdelta X^{M_2}(\sigma_2)\FF^\ha_{M_1,M_2}(\sigma_1,\sigma_2)[X] 
\cr&=&\frac12\int_{\Sigma_p}\,   \iota_{\Fdelta X^{M_1}(\sigma) \Fwedge  \Fdelta X^{M_2}(\sigma)}\rd A^\ha_{[p+1]}(X(\sigma))
\cr&&+\frac{\ii}2 f^\ha{}_{\hb_1\hb_2}\int_{\Sigma_p}\iota_{\Fdelta X^{M_1}(\sigma_1)} A^{\hb_1}_{[p+1]}(X(\sigma_1))\Fwedge\int_{\Sigma_p}\iota_{\Fdelta X^{M_2}(\sigma_2)} A^{\hb_2}_{[p+1]}(X(\sigma_2))\,.
\end{eqnarray}
Unlike the abelian case, we cannot define a local $(p+2)$-form gauge field strength since $\FF^\ha_{M_1,M_2}(\sigma_1,\sigma_2)[X]$ is typically a bi-local quantity from the perspective of the target space.  

\section{A toy matrix model}
In Sec.III, it is shown that the target space $p$-form gauge symmetry can be regarded as a specific construction of the generic 0-form gauge symmetry in the functional space $\{X^M(\sigma)\}$ of $p$-brane configurations. For $p>0$, the theory is complicated since one need to deal with infinite dimensional functional space $\{X^M(\sigma)\}$. In this section, we develop a much simpler toy model in which the infinite dimensional functional space $\{X^M(\sigma)\}$ is replaced by a finite dimensional matrix space.
 
\vspace{2ex}
\noindent \underline{\bf{Matrix string model}}
\vspace{2ex}

For simplicity, let us consider the $p=1$ case firstly. The basic idea is to replace the continuous spatial coordinate $\sigma$ on the string worldsheet by the discrete indices of a $2\times2$ matrix.  That is
\begin{eqnarray}
\sigma&\leftrightarrow& ^{a}{}_{b}\,,~~~~~~a,b=1,2\,.
\end{eqnarray}
For a field $\phi(\sigma)$ on string, the infinite dimensional configuration space $\{\phi(\sigma)\}$ is replaced by the space $\cM^{(2)}$ of $2\times2$ matrices. 
Under the equal time mode expansion, $\phi(\sigma)$ will be expanded by the  
``eigenfunctions'' $(\Eif_{\kappa})^{a}{}_{b}$
\begin{eqnarray} 
\Eif_{0}&=&\!\bm1\,,~~~\Eif_{\pm}=\hat\sigma_{\pm}=\frac12(\hat\sigma_1\pm\ii\hat\sigma_2)\,,~~~~~ \Eif_{\dZ}=\hat\sigma_3 \,,
\end{eqnarray}
where $\hat\sigma_i$'s are the Pauli matrices.
For example, $X^{M}(\sigma)$ becomes 
\begin{eqnarray} 
X^{M}(\sigma)&\to&\!x^M\Eif_{0}+l_s\sum_{\kappa=\pm}\chi^{M\kappa}\Eif_{\kappa} +l_s^2w^{M}\Eif_{\dZ}
=\left(
\begin{array}{cc}
x^{M}+l_s^2 w^{M} & l_s\chi^{M,+} \\
l_s\chi^{M,-} & x^{M}-l_s^2 w^{M} \\
\end{array}
\right)\,,
\end{eqnarray}
where $l_{s}=l_1=T_{1}^{-\frac12}$ is the string length parameter. 

The products of functions on string are mapped to
\begin{eqnarray}\label{def1}
\phi(\sigma)\psi(\sigma') &\leftrightarrow& \phi^{a}{}_{b}\psi^{a'}{}_{b'}\,,
\cr
\phi(\sigma)\psi(\sigma)=(\phi\psi)(\sigma)&\leftrightarrow& (\phi\psi)^{a}{}_{b}=\phi^{a}{}_{c}\psi^{c}{}_{b}\,. 
\end{eqnarray}
The spatial integration on the string worldsheet becomes the symmetric trace
\begin{eqnarray}\label{def2}
\int\rd\sigma\, \phi_1(\sigma)\cdots\phi_m(\sigma) &\leftrightarrow& {\rm Tr}_{\rm sym}(\phi_1\cdots\phi_m)=\frac1{m!}{\rm Tr}(\phi_1\cdots\phi_m+{\rm permutations})\,.
\end{eqnarray}}
Especially,  
\begin{eqnarray}
V_1=\int\rd\sigma\,&\leftrightarrow& {\rm Tr}\bm1=2\,. 
\end{eqnarray}
Although the matrix product is non-commutative, we don't need to worry about it inside the symmetric trace.

The inner product of the functions on string becomes
\begin{eqnarray}
\langle\phi ,\psi \rangle=\frac1{V_1}\int\rd\sigma\,\phi^{\dagger}(\sigma)\psi(\sigma) &\leftrightarrow& \frac12{\rm Tr}(\phi^{\dagger}\psi)\,. 
\end{eqnarray}
Thus the ``eigenfunctions'' $\Eif_{\kappa}$ satisfies the orthonormality condition
\begin{eqnarray}
\langle\Eif_{\tilde\kappa},\Eif_\kappa\rangle=\frac12{\rm Tr}[\Eif^{\dagger}_{\tilde\kappa}(\sigma)\Eif_\kappa(\sigma)]=\delta_{\tilde\kappa \kappa}\,. 
\end{eqnarray}
Summing over these ``eigenfunctions'', we get the $\delta$-function of this matrix string model
\begin{eqnarray}\label{def3}
\delta(\sigma;\sigma')=\sum_{\kappa}\Eif_{\kappa}(\sigma)\Eif^{\dagger}_{\kappa}(\sigma')&\leftrightarrow&
\delta^{a;a'}{}_{b;b'}=\frac12\left[\delta^{a}{}_{b}\delta^{a'}{}_{b'} +(\hat\bmsigma^{m})^{a}{}_{b}(\hat\bmsigma_{m})^{a'}{}_{b'}\right]
=\delta^{a}{}_{b'}\delta^{a'}{}_{b} \,. 
\end{eqnarray}
One can easily confirm that the fundamental property of the $\delta$-function is satisfied 
\begin{eqnarray}
\label{delta1}
\int\rd \sigma'\,\delta(\sigma;\sigma')\psi(\sigma')&\leftrightarrow&
\delta^{a;a'}{}_{b;c'}\psi^{c'}{}_{a'}=\delta^{a}{}_{c'}\delta^{a'}{}_{b}\psi^{c'}{}_{a'}=\psi^{a}{}_{b}\leftrightarrow\psi(\sigma)\,. 
\end{eqnarray}
It is also obvious that 
\begin{eqnarray}\label{delta2}
\frac{\delta \phi(\sigma)}{\delta\phi(\sigma')}&\leftrightarrow& \frac{\partial \phi^{a}{}_{b}}{\partial\phi^{a'}{}_{b'}}=\delta^{a}{}_{a'}\delta^{b'}{}_{b} \leftrightarrow \delta(\sigma;\sigma')\,.
\end{eqnarray}

The spatial derivative operator on the worldsheet is realized by
\begin{eqnarray}\label{def4}
\partial_{\sigma}\phi(\sigma)&\leftrightarrow&l_s^{-1}[\hat\sigma_3,\phi]\,.
\end{eqnarray}
By construction, it satisfies the Leibniz rule
\begin{eqnarray}\label{d1}
\partial_{\sigma}(\phi\psi)&\leftrightarrow& l_s^{-1}[\hat\sigma_3,\phi\psi]=l_s^{-1}[\hat\sigma_3,\phi]\psi+l_s^{-1}\phi[\hat\sigma_3,\psi]
\leftrightarrow\partial_{\sigma}\phi\psi+\phi\partial_{\sigma}\psi\,. 
\end{eqnarray} 
Integrating over the string, we get
\begin{eqnarray}\label{d2}
\int\rd\sigma\,\partial_\sigma\phi(\sigma) &\leftrightarrow&  l_s^{-1}{\rm Tr}[\hat\bmsigma_3,\phi]=0\,,
\end{eqnarray} 
and
\begin{eqnarray}\label{delta3}
\int\rd\sigma'\partial_{\sigma'}\delta(\sigma;\sigma')\phi(\sigma') &\leftrightarrow&  
l_s^{-1}[(\hat\bmsigma_{3})^{b'}{}_{c'}\delta^{a;c'}{}_{b;a'} -\delta^{a;b'}{}_{b;c'}(\hat\bmsigma_{3})^{c'}{}_{a'}]\phi^{a'}{}_{b'} \cr 
\cr&=&l_s^{-1}[(\hat\bmsigma_{3})^{b'}{}_{b}\phi^{a}{}_{b'} -\phi^{a'}{}_{b} (\hat\bmsigma_{3})^{a}{}_{a'}]
=-l_s^{-1}[\hat\bmsigma_{3},\phi]^{a}{}_{b}\leftrightarrow
-\partial_{\sigma}\phi(\sigma)\,.
\end{eqnarray}
 
Given a target space field $\Phi(x)$, its pullback on the matrix worldsheet is defined by the formal Taylor expansion
\begin{eqnarray}\label{def5}
\Phi(X)&=&\sum_{n=0}^{\infty}\frac{1}{n!}\partial_{N_n}\cdots\partial_{N_2}\partial_{N_1}\Phi(x) (X^{N_1})^{a}{}_{a_2} (X^{N_2})^{a_2}{}_{a_3}\cdots (X^{N_n})^{a_{n}}{}_{b}
\,.
\end{eqnarray}
Now, together with the definitions (\ref{def1}, \ref{def2}, \ref{def3}, \ref{def4}),  
all the $\sigma$-integrated expression appeared in the original string model can be transplanted to our matrix models on $\cM^{(2)}$. Due to the properties (\ref{delta1}, \ref{delta2}, \ref{d1}, \ref{d2}, \ref{delta3}), the discussions in Sec.III are directly inherited by the present matrix model for $p=1$. For example, the target space 2-form gauge field $B_{MN}(x)$ is used to construct the 1-form gauge field $\FA^{\ha}[X]$ in the matrix string space 
\begin{eqnarray}
\FA^{\ha}[X]=\int_{\Sigma_1}\rd\sigma\,\Fdelta X^{M}(\sigma)B^{\ha}_{MN}(X)\partial_{\sigma}X^{N}(\sigma)
 &\leftrightarrow& {\rm Tr}_{\rm sym}\Big[\rd X^{M}B^{\ha}_{MN}(X)\partial_{\sigma}X^{N}\Big]\,.
\end{eqnarray}
The target space 1-form gauge transformation is realized as the 0-form gauge transformation in the matrix string space
\begin{eqnarray}
u^\ha[X]=\int_{\Sigma_1}\rd\sigma\, u^\ha_{N}(X)\partial_{ \sigma}X^{N}(\sigma)
&\leftrightarrow& {\rm Tr}\Big(u^{\ha}_{N}(X)\partial_{\sigma}X^{N}\Big)\,.
\end{eqnarray}  
The same discussion as in Sec.III implies that 
\begin{eqnarray}
\delta_{\cG} B^\ha_{MN}&=&-2 \partial_{[M}u^\ha_{N]}-\ii f^\ha{}_{\hb\hc} B^\hb_{MN}{\rm Tr}\Big(u^{\hc}_{N_1}(X)\partial_{\sigma}X^{N_1}\Big)\,. 
\end{eqnarray}

\vspace{2ex}
\noindent \underline{\bf{Matrix $p$-brane model}}
\vspace{2ex}

For generic $p$-branes, one just need to consider the tensor product of $p$-copies $2\times 2$ matrix 
\begin{eqnarray}
\cM^{(2,p)}=\cM^{(2)}\otimes\cdots\otimes\cM^{(2)}\,.
\end{eqnarray} 
Thus
\begin{eqnarray}
X^M(\sigma) &\leftrightarrow&  (X^{M})^{a_1\cdots a_p}{}_{b_1\cdots b_p}\,,~~~~~~~~a_i,b_i=1,2\,, 
\end{eqnarray}
where the $i$-th worldvolume spatial coordinate $\sigma^i$ is identified with the indices ${}^{a_i}{}_{b_i}$.

All the definitions (\ref{def1}, \ref{def2}, \ref{def3}, \ref{def4}, \ref{def5}) above can be straightforwardly generalized to $p>1$ cases. Especially, the derivative operators are realized as  
\begin{eqnarray}
\partial_{\sigma^1}\phi(\sigma) &\leftrightarrow& [\hat\sigma_3\otimes\bm1\otimes\bm1\otimes\cdots \otimes\bm1 ,\phi]\,,
\cr \partial_{\sigma^2}\phi(\sigma) &\leftrightarrow& [\bm1\otimes\hat\sigma_3\otimes\bm1\otimes\cdots \otimes\bm1 ,\phi]\,,
\\ &\vdots& \nn
\end{eqnarray}
such that the property
\begin{eqnarray}
\partial_{\sigma^i}\partial_{\sigma^j}\phi(\sigma)=\partial_{\sigma^j}\partial_{\sigma^i}\phi(\sigma)
\end{eqnarray} 
is satisfied. Besides, one can also easily generalize the properties (\ref{delta1}, \ref{delta2}, \ref{d1}, \ref{d2}, \ref{delta3}) for $p>1$. As a result, the discussions of $p$-form gauge symmetry in Sec.III are fully inherited by the matrix model based on $\cM^{(2,p)}$.

\section{Summary and Discussion}
By considering the phase ambiguity of the $p$-brane wave functional, it is shown that the target space $p$-form gauge symmetry can be regarded as a special construction of the generic 0-form gauge symmetry in the functional space $\{X^{M}(\sigma)\}$ of $p$-brane configurations. These discussions are valid both for the abelian and non-abelian gauge groups. Furthermore, we develop a toy model for $p$-form gauge symmetry by replacing the infinite dimensional $p$-brane configuration space with a finite dimensional matrix space. 

To construct a gauge invariant action for $p$-form gauge theory, it is also natural to start from the $p$-brane configuration space. We notice that the target space metric induces a natural metric on $\{X^{M}(\sigma)\}$ 
\begin{eqnarray}
\rd s^2_{[X(\sigma)]}&=&\int\rd^p \sigma  \int\rd^p \sigma'  \, \FG_{M,N}(\sigma,\sigma')[X]\Fdelta X^{M}(\sigma)\Fdelta X^{N}(\sigma')
=\int\rd^p \sigma\,G_{MN}(X)\Fdelta X^{M}(\sigma)\Fdelta X^{N}(\sigma)\,.
\end{eqnarray}
Then one can naively propose a Yang-Mills type of Lagrangian on $\{X^{M}(\sigma)\}$  
\begin{eqnarray}
\cL[X(\sigma)]&=&\int\rd^p \sigma_1 \int\!\!\rd^p \sigma_2 \int\!\!\rd^p \sigma_1'\!\int\!\!\rd^p \sigma_2'\, \FG^{M_1,M_1'}(\sigma_1,\sigma_1')[X]\FG^{M_2,M_2'}(\sigma_2,\sigma_2')[X] 
\cr&&~~~~~~~~~~~~~~~~~~~~~~~~~~~~~~~~~~~\times{\rm Tr}_{\cG}\left\{\FF_{M_1,M_2}(\sigma_1,\sigma_2)[X]\FF_{M_1',M_2'}(\sigma_1',\sigma_2')[X]\right\}\,.
\end{eqnarray}
The action is obtained by integrating over the $p$-brane configurations   
\begin{eqnarray}
S_{\{X(\sigma)\}}&=&\int[\cD X(\sigma)]\,\cL[X(\sigma)]\,. 
\end{eqnarray} 
In terms of the mode expansion
\begin{eqnarray}
X^{M}(\sigma)=x^{M}+\sum_{k_i} a^{M}_{k_i}\ep^{\ii k_i\sigma^i}\,,
\end{eqnarray}
the integration measure is converted to 
\begin{eqnarray}
\int[\cD X(\sigma)]=\int\rd x\prod_{k_i} \rd a_{k_i} \,.
\end{eqnarray} 
To derive an effective target space Lagrangian, we need to  integrate out the tower of oscillating modes 
\begin{eqnarray}
\cL_{\rm eff}(x)=\int[\prod_{k_i}\rd a_{k_i}]\,\cL[X(\sigma)] \,. 
\end{eqnarray}
Of course, it is rather difficult to perform the infinite dimensional integration in the original $p$-brane model in Sec.III.  On the other hand, the number of oscillating modes is finite in the matrix model introduced in Sec.IV. Thus it would be much easier to get a target space $p$-form gauge theory in the matrix model approach.

\vspace*{3.0ex}
\begin{acknowledgments}
\paragraph*{Acknowledgments.} 
The authors thank Bo-Han Li, Jian-Xin Lu, Hong L\"u and Jun-Bao Wu for useful conversations.
This work is supported by National Natural Science Foundation of China(Grants No. 12275217, No. 12247103).
\end{acknowledgments}

\bibliographystyle{unsrturl}


\end{document}